# Theoretical, Numerical, and Experimental Evidence of Superluminal Electromagnetic and Gravitational Fields Generated in the Nearfield of Dipole Sources

(Presented at Days on Diffraction Conference, St. Petersburg, Russia, June 2005)

**William D. Walker**
Örebro University, Department of Technology, Sweden
Research papers [1]
william.walker@tech.oru.se

## Abstract

Theoretical and numerical wave propagation analysis of an oscillating electric dipole is presented. The results show that upon creation at the source, both the longitudinal electric and transverse magnetic fields propagate superluminally and reduce to the speed of light as they propagate about one wavelength from the source. In contrast, the transverse electric field is shown to be created about 1/4 wavelength outside the source and launches superluminal fields both towards and away from the source which reduce to the speed of light as the field propagates about one wavelength from the source. An experiment using simple dipole antennas is shown to verify the predicted superluminal transverse electric field behavior. In addition, it is shown that the fields generated by a gravitational source propagate superluminally and can be modeled using quadrapole electrodynamic theory. The phase speed, group speed, and information speed of these systems are compared and shown to differ. Provided the noise of a signal is small and the modulation method is known, it is shown that the information speed can be approximately the same as the superluminal group speed. According to relativity theory, it is known that between moving reference frames, superluminal signals can propagate backwards in time enabling violations of causality. Several explanations are presented which may resolve this dilemma.

## Introduction

The electromagnetic fields generated by an oscillating electric dipole have been theoretically studied by many researchers using Maxwell's equations and are known to yield the following well-known results (MKS units):

## System differential equation

**Fig. 1**

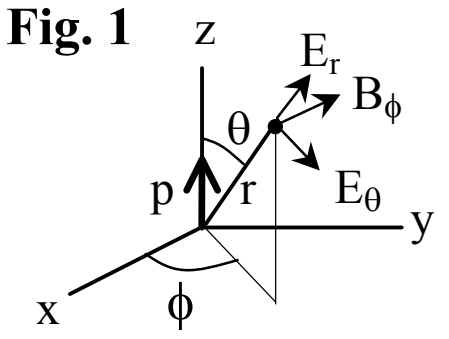

System PDE

$$\nabla^2 V - \frac{1}{c^2}\frac{\partial^2 V}{\partial t^2} = \frac{-\rho}{\varepsilon_o} \quad (1)$$

Variable definitions
- $E_r$ = Radial electric field
- $E_\theta$ = Transverse electric field
- $B_\phi$ = Transverse magnetic field
- $V$ = Scalar potential
- $\rho$ = Charge density
- $\varepsilon_o$ = Free-space permittivity
- $\nabla^2$ = Laplacian
- c = Speed of light
- t = Time
- p = Dipole (q d)
- $\omega$ = Angular frequency
- k = Wave number

## Field analysis

Solving the homogeneous equation (Eq. 1) for a dipole source yields [2, 3, 4, 5, 6]:

$$V = N\,Cos(\theta)\left[\frac{1}{kr} + \frac{i}{(kr)^2}\right]e^{i(kr-\omega t)} \qquad \text{where: } N = \frac{pk^2}{4\pi\varepsilon_o} \quad (2)$$

The fields can then be calculated using the following relations [6]:

$$B = \frac{\omega}{c^2}(r \times \nabla V) \qquad E = \frac{ic^2}{\omega}(\nabla \times B) \quad (3)$$

yielding: $$E_r = \frac{pCos(\theta)}{2\pi\varepsilon_o r^3}\left[1 - i(kr)\right]e^{i(kr-\omega t)}$$

$$E_\theta = \frac{pSin(\theta)}{4\pi\varepsilon_o r^3}\left[\left\{1-(kr)^2\right\} - i(kr)\right]e^{i(kr-\omega t)} \qquad B_\phi = \frac{\omega p Sin(\theta)}{4\pi\varepsilon_o c^2 r^2}\left[-kr - i\right]e^{i(kr-\omega t)} \quad (4)$$

## Phase analysis

The general form of the electromagnetic fields generated by a dipole is:

$$Field \propto (x + iy)\cdot e^{i[kr-\omega t]}$$

If the source is modeled as $Cos(\omega t)$, the resultant generated field is:

$$Field \propto Mag \cdot Cos[\{kr + ph\} - \omega t] = Mag \cdot Cos(\theta - \omega t)$$

where: $\mathrm{Mag} = \sqrt{\mathrm{x}^2 + y^2}$

It should be noted that the formula describing the phase is dependent on the quadrant of the complex vector.

$$\theta_1 = kr + Tan^{-1}\left(\frac{y}{x}\right) \qquad \theta_2 = kr - Cos^{-1}\left(\frac{x}{\sqrt{x^2+y^2}}\right) \tag{5}$$

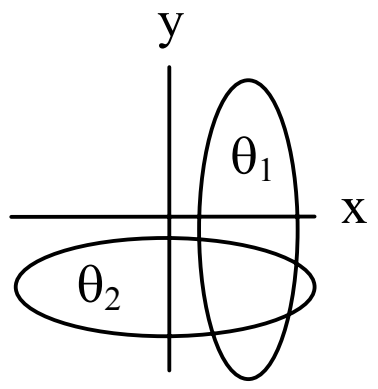

## Phase speed analysis

Phase speed can be defined as the speed at which a wave composed of one frequency propagates. The phase speed ($c_{ph}$) of an oscillating field of the form $Sin(\omega t - kr)$, in which $k = k(\omega, r)$, can be determined by setting the phase part of the field to zero, differentiating the resultant equation, and solving for $\partial r/\partial t$.

$$\frac{\partial}{\partial t}(\omega t - kr) = 0 \qquad \therefore \omega - k\frac{\partial r}{\partial t} - r\frac{\partial k}{\partial r}\frac{\partial r}{\partial t} = 0 \qquad \therefore c_{ph} = \frac{\partial r}{\partial t} = \frac{\omega}{k + r\dfrac{\partial k}{\partial r}} \tag{6}$$

Differentiating $\theta \equiv -kr$ with respect to r yields: $\dfrac{\partial \theta}{\partial r} = -k - r\dfrac{\partial k}{\partial r}$ (7)

Combining these results and inserting the far-field wave number ($k = \omega/c_o$) yields:

$$c_{ph} = -\omega \Big/ \frac{\partial \theta}{\partial r} = -c_o k \Big/ \frac{\partial \theta}{\partial r} \tag{8}$$

## Group speed analysis

The group speed of an oscillating field of the form: $Sin(\omega t - kr)$, in which $k = k(\omega, r)$, can be calculated by considering two Fourier components of a wave group [7]: $Sin(\omega_1 t - k_1 r) + Sin(\omega_2 t - k_2 r) = Sin(\Delta\omega t - \Delta k r)\ Sin(\omega t - kr)$ (9) in which: $\Delta\omega = \dfrac{\omega_1 - \omega_2}{2}$, $\Delta k = \dfrac{k_1 - k_2}{2}$, $\omega = \dfrac{\omega_1 + \omega_2}{2}$, $k = \dfrac{k_1 + k_2}{2}$

The group speed ($c_g$) can then be determined by setting the phase part of the modulation component of the field to zero, differentiating the resultant equation, and solving for $\partial r/\partial t$:

$$\frac{\partial}{\partial t}(\Delta\omega t - \Delta k r) = 0 \qquad \therefore \Delta\omega - \Delta k\frac{\partial r}{\partial t} - r\frac{\partial \Delta k}{\partial r}\frac{\partial r}{\partial t} = 0 \qquad \therefore c_g = \frac{\partial r}{\partial t} = \frac{\Delta\omega}{\Delta k + r\dfrac{\partial \Delta k}{\partial r}} \tag{10}$$

Differentiating $\Delta\theta \equiv -\Delta k r$ with respect to r yields: $\dfrac{\partial \Delta\theta}{\partial r} = -\Delta k - r\dfrac{\partial \Delta k}{\partial r}$ (11)

Combining these results and using the far-field wave number ($k = \omega/c_o$) yields:

$$c_g = -\Delta\omega \Big/ \frac{\partial \Delta\theta}{\partial r} = -\left[\frac{\partial}{\partial r}\frac{\Delta\theta}{\Delta\omega}\right]^{-1} \qquad \therefore c_g \underset{\lim \frac{\Delta\theta}{\Delta\omega} small}{=} -\left[\frac{\partial^2\theta}{\partial r \partial\omega}\right]^{-1} = -c_o\left[\frac{\partial^2\theta}{\partial r \partial k}\right]^{-1} \tag{12}$$

It should be noted that other derivations of the above phase and group speed relations are available in previous publications by the author [8, 9, 10, 11, 12, 13] and in the following well-known reference [14].

In addition, in order for the group speed to be valid, a signal should not distort as it propagates. It is known from electronic signal theory that in order to minimize signal distortion, the phase vs. frequency curve must be approximately linear over the bandwidth of the signal and the amplitude vs. frequency curve must be approximately constant over the bandwidth of the signal [15]. It is shown below that the amplitude vs. frequency curve can even be approximately linear over the bandwidth of the signal, provided the ratio of the slope of the curve to the signal amplitude is small. Assuming that the amplitude vs. frequency curve is increasing and approximately linear over the bandwidth of a modulated carrier signal, each signal magnitude (A) Fourier component ($w_m$) will be increased by (u) and the Fourier component symmetric about the carrier ($w_c$) will be reduced by (u):

$$\frac{1}{2}\left[(A-u)\ Sin(w_c t - w_m t) + (A+u)\ Sin(w_c t + w_m t)\right]$$
$$= A \cdot Cos(w_m t) Sin(w_c t) + u \cdot Sin(w_m t) Cos(w_c t) \quad (13)$$

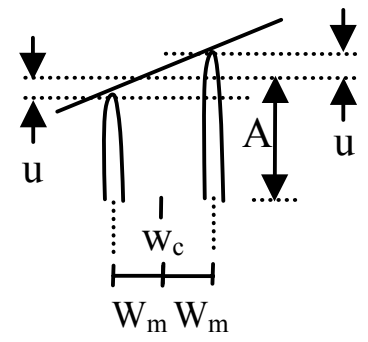

The two Fourier components form an amplitude modulated signal where the magnitude of the carrier is:

$$\sqrt{A^2 \cdot Cos^2(w_m t) + u^2 \cdot Sin^2(w_m t)} \approx A \cdot Cos(w_m t) \quad (14)$$

It should be noted that distortions to the magnitude are minimal provided: $u^2/A^2 << 1$

where (u) can be approximated using the derivative relation: $u = \frac{\Delta k}{2} \frac{\partial A}{\partial k}$, (15)

provided the amplitude vs. frequency curve is approximately linear over the bandwidth of the signal. In addition, it should be noted that phase speed and group speed can also be determined from two different frequency components ($\omega_1$, $\omega_1$) using relations (Eq. 8, 12):

$$c_{ph} = -\omega_c \Big/ \frac{\partial \theta}{\partial r} \qquad c_g = -\Delta\omega \Big/ \frac{\partial \Delta\theta}{\partial r} \qquad \text{Given} \quad \Delta\omega = \frac{\omega_2 - \omega_1}{2} = \omega_m , \quad \omega_c = \frac{\omega_1 + \omega_2}{2}$$

yields:
$$c_{ph} = \frac{-\omega_c}{\frac{\partial}{\partial r}\left(\frac{\theta_2 + \theta_1}{2}\right)} \qquad c_g = \frac{-\omega_m}{\frac{\partial}{\partial r}\left(\frac{\theta_2 - \theta_1}{2}\right)} \quad (16)$$

Given two different frequencies, plots of the phase speed and group speed can then be determined for each field component by inserting the corresponding phase relation (Eq. 17, 20, 23). It should be noted that these relations yield the same results as (Eq. 8, 12) provided the phase and magnitude vs. frequency curves are approximately linear over the bandwidth of the signal.

## <u>Wave propagation analysis of near-field electric dipole EM fields</u>

To determine how the EM fields propagate in an electric dipole system, one can apply the above phase and group speed relations (Eq. 8, 12) to the known theoretical solution of an electric dipole (Eq. 4).

## Radial electric field (Er) solution

$$y = -kr \qquad x = 1$$

$$\theta = kr - Tan^{-1}(kr) \underset{kr<<1}{\approx} -\frac{1}{3}(kr)^3 \tag{17}$$

$$c_{ph} = c_o\left(1 + \frac{1}{(kr)^2}\right) \underset{kr<<1}{\approx} \frac{c_o}{(kr)^2} \underset{kr>>1}{\approx} c_o \tag{18}$$

$$c_g = \frac{c_o\left(1+(kr)^2\right)^2}{3(kr)^2 + (kr)^4} \underset{kr<<1}{\approx} \frac{c_{ph}}{3} \underset{kr>>1}{\approx} c_o \tag{19}$$

$E_r$ Phase vs. kr
**Fig. 2**

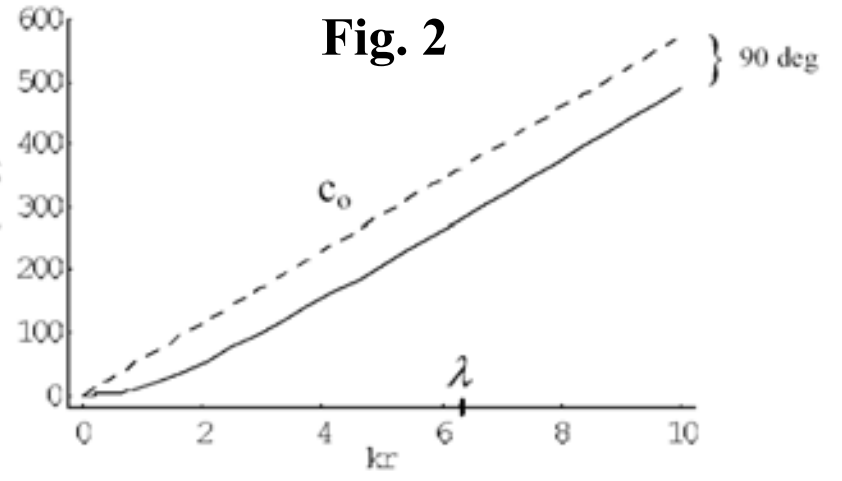

$E_r$ $c_{ph}/c_o$ vs. kr
**Fig. 3**

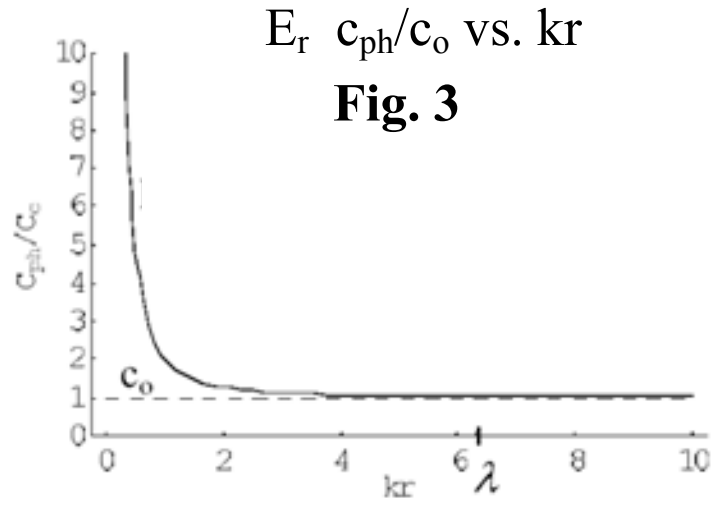

$E_r$ $c_g/c_o$ vs. kr
**Fig. 4**

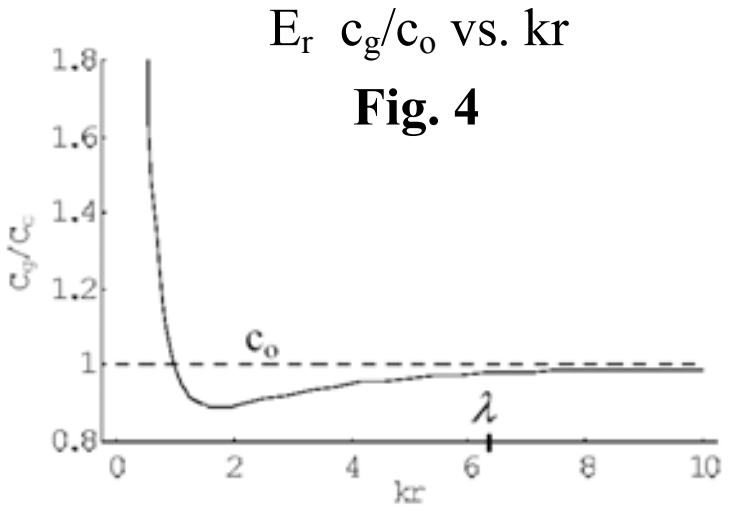

## Transverse electric field ($E_\theta$) solution

$$y = -kr \qquad x = 1-(kr)^2$$

$$\theta = kr - Cos^{-1}\left(\frac{1-(kr)^2}{\sqrt{1-(kr)^2+(kr)^4}}\right) \tag{20}$$

$$c_{ph} = c_o\left(\frac{1-(kr)^2+(kr)^4}{-2(kr)^2+(kr)^4}\right) \tag{21}$$

$$c_g = \frac{c_o\left(1-(kr)^2+(kr)^4\right)^2}{-6(kr)^2+7(kr)^4-(kr)^6+(kr)^8} \tag{22}$$

$E_\theta$ Phase vs. kr
**Fig. 5**

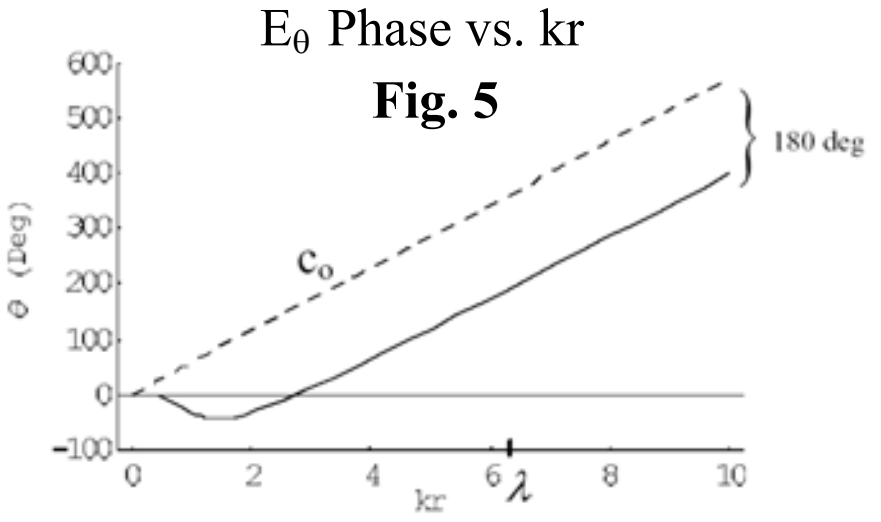

$E_\theta$ $c_{ph}/c_o$ vs. kr
**Fig. 6**

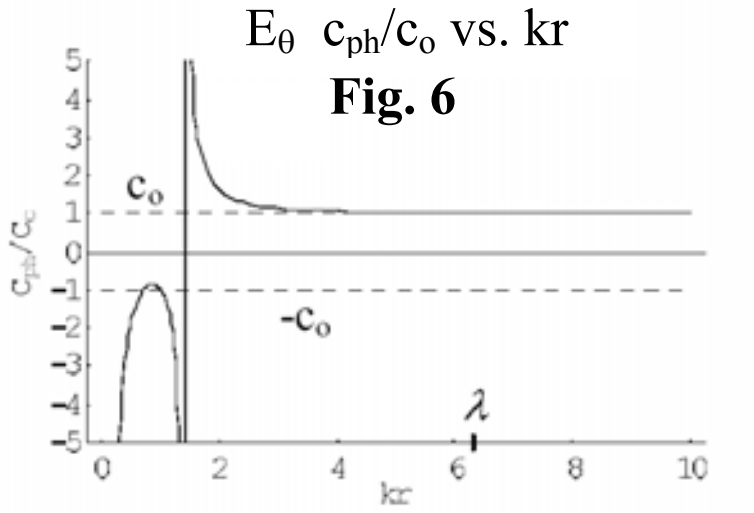

$E_\theta$ $c_g/c_o$ vs. kr
**Fig. 7**

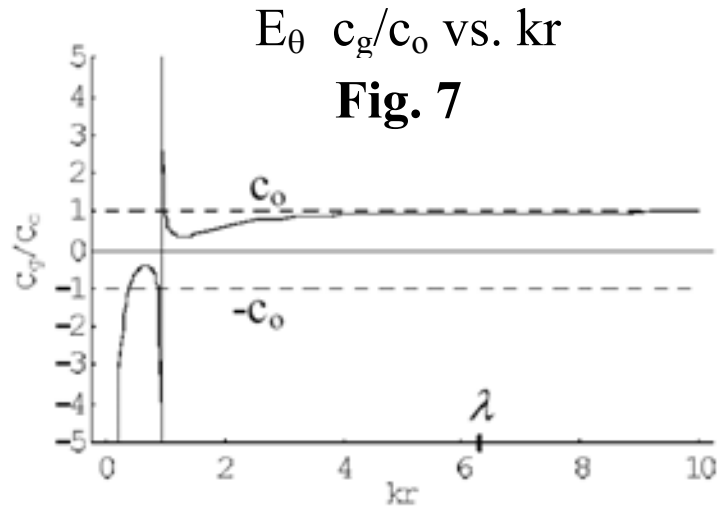

## Transverse magnetic field ($B_\phi$) solution

$$y = -1 \qquad x = -kr$$

$$\theta = kr - Cos^{-1}\left(\frac{-kr}{\sqrt{1+(kr)^2}}\right) \tag{23}$$

$$c_{ph} = c_o\left(1 + \frac{1}{(kr)^2}\right) \tag{24}$$

$$c_g = \frac{c_o\left(1+(kr)^2\right)^2}{3(kr)^2 + (kr)^4} \tag{25}$$

$B_\phi$ Phase vs. kr
**Fig. 8**

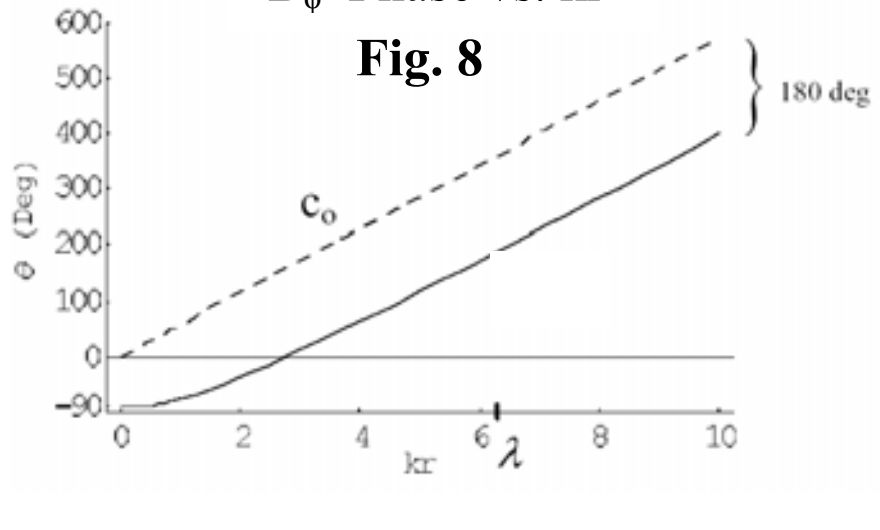

$B_\phi$ $c_{ph}/c_o$ vs. kr
**Fig. 9**

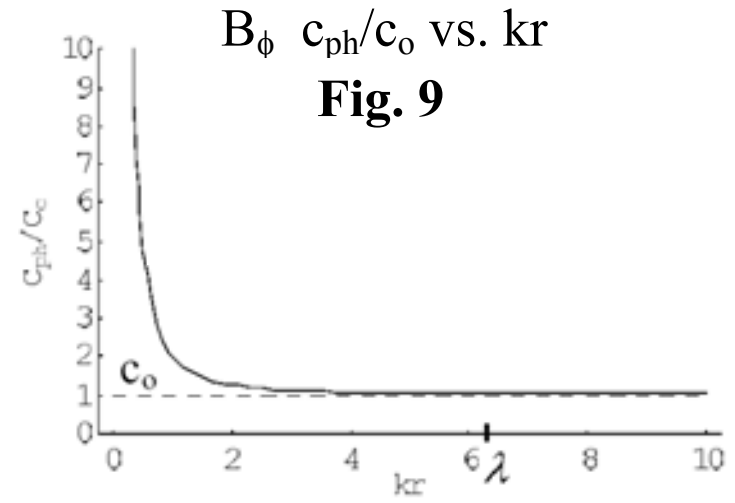

$B_\phi$ $c_g/c_o$ vs. kr
**Fig. 10**

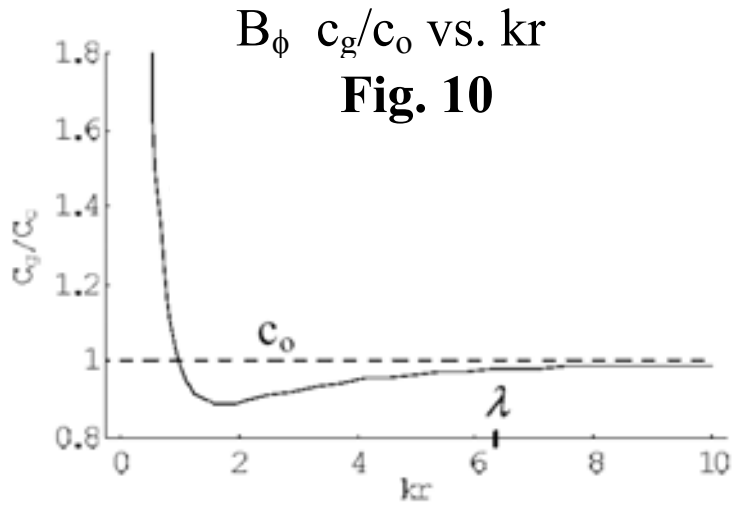

The above results (p.4) were originally published by the author in 1999 [10], but the propagation of the longitudinal electric field and gravitational field next to a source was published earlier by the author in 1997 [12, 13]. It should be noted that after these dates similar results have been published by other authors [16, 17].

## Numerical verification of field component propagation

To verify the predicted wave propagation effects, a numerical simulation was performed. The simulation consisted of extracting the transfer function from the various field components of the known dipole solution (Eq. 4) and then inverse Fourier transforming ($FT^{-1}$) the Fourier transform (FT) of a given signal multiplied by the dipole transfer function [18]: Result Sig = $FT^{-1}$ [ FT [Signal] x G ], where "Result Sig" is the resultant propagating signal as a function of time, and "G" is the transfer function of the wave propagation system. A simple amplituded modulated signal (300MHz carrier, 20MHz modulation), generated by adding together two sinusoidal oscillations (280MHz and 320MHz), was applied as an input signal. The simulations yielded propagating signal-versus-time animations as a function of distance (r) from the source. To compare the resulting simulations to theoretical expectations, a propagating modulation envelope of the AM signal was superimposed: Mod = $\cos[w_m t-(\theta_2-\theta_1)/2]$ where $\theta_1$ and $\theta_2$ are the theoretically expected phase shifts for the 2 frequencies used to create the AM signal (Eq. 17, 20, 23). The results below show a very good match between theory and numerical simulation.

**Fig. 11**

**Resultant $E_r$ vs. time animation plots as a function of distance (r) from source**

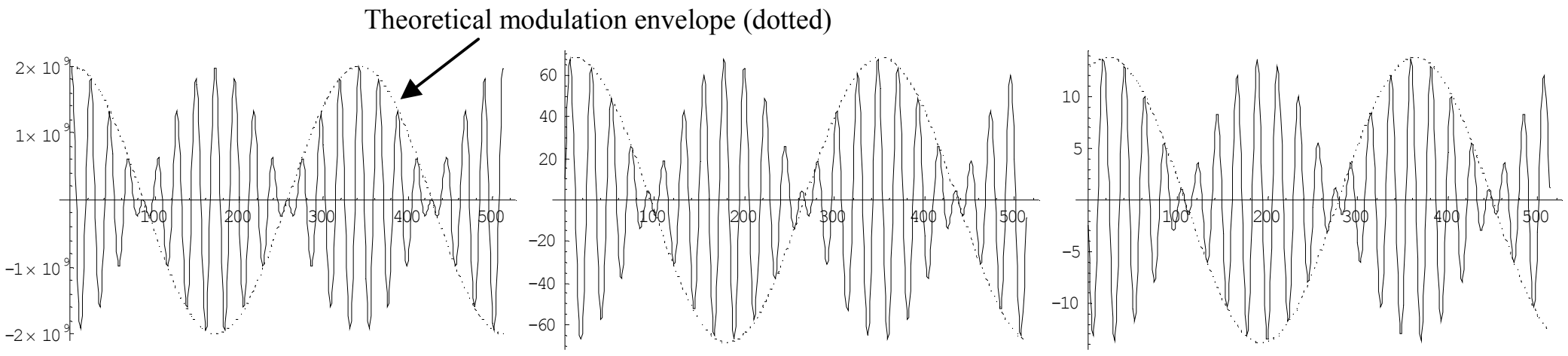

(1/25)$\lambda_c$ (12/25)$\lambda_c$ (25/25)$\lambda_c$

**Fig. 12**

**Resultant $E_\theta$ vs. time animation plots as a function of distance (r) from source**

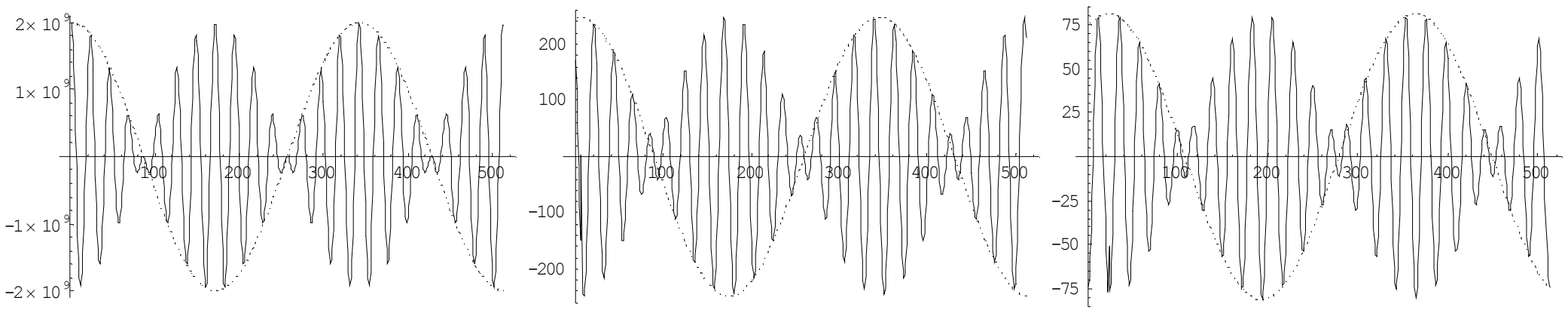

(1/25)$\lambda_c$ (8/25)$\lambda_c$ (25/25)$\lambda_c$

**Fig. 13**

**Resultant $H_\phi$ vs. time animation plots as a function of distance (r) from source**

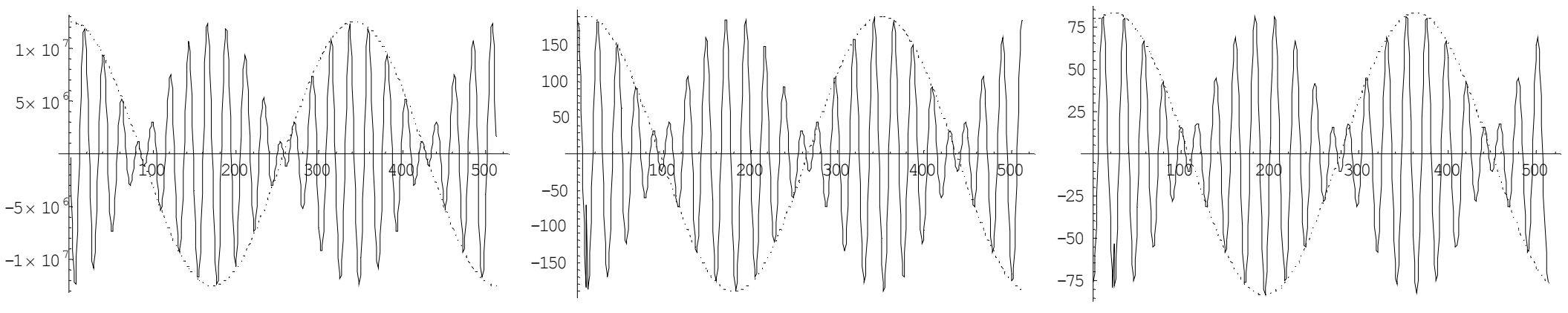

(1/25)$\lambda_c$ (12/25)$\lambda_c$ (25/25)$\lambda_c$

**Fig. 14**

**Mathematica code for simulation of $E_\theta$ wave propagation**

```
f1=280*10^6; f2=320*10^6; fc=(f1+f2)/2; fm=(f2-f1)/2; w1=2*Pi*f1; w2=2*Pi*f2; T=1/fm;
x=Cos[w1*t]+Cos[w2*t];      (* Input signal *)
n=1024; Cycle=1.5; ts=Cycle*T/n; fs=1/ts; fn=fs/2; v=N[Table[x,{t,0,Cycle*T,Cycle*T/n}]];
y=Fourier[v]*1.4; k=2*Pi/L; L=c/f; c=3*10^8; u=N[Table[k,{f,0,fs,fs/n}]];
h=y*(1-(u*r)*(u*r)-I*(u*r))/r^3*Exp[I*u*r];      (* FT [Signal] x G *)
l=Take[h,n/2]; <<Graphics`Animation` <<Graphics`MultipleListPlot` L1=c/f1; L2=c/f2; k1=2*Pi/L1;
k2=2*Pi/L2; km=(k2-k1)/2; kc=(k2+k1)/2; wm=2*Pi*fm; wc=2*Pi*fc;
ph1=k1*r-ArcCos[(1-(-k1*r)^2)/Sqrt[(1-(k1*r)^2)+(k1*r)^4]];     (* Eθ phase relation for f1 *)
ph2=k2*r-ArcCos[(1-(-k2*r)^2)/Sqrt[(1-(k2*r)^2)+(k2*r)^4]];     (* Eθ phase relation for f2 *)
tn=N[Table[t,{t,0,Cycle*T,Cycle*T*2/n}]];
Md=2*Abs[(1-(wc/c*r)*(wc/c*r)-I*(wc/c*r))/r^3]*Cos[wm*tn-(ph2-ph1)/2];  (* Theoretical cg plot *)
Animate[MultipleListPlot[Re[InverseFourier[l]],Md,PlotJoined->True,
   SymbolShape\[Rule]None],{r,0.001,c/fc,c/fc/25}]
```

Er animations were generated using the above code and substituting the following known Er relations

```
h=y*(1-I*(u*r))/r^3*Exp[I*u*r];   (* FT [Signal] x G *)
ph1=k1*r-ArcTan[k1*r];   (* Er phase relation for f1 *)
ph2=k2*r-ArcTan[k2*r];   (* Er phase relation for f2 *)
Md=2*Abs[(1-I*(wc/c*r))/r^3]*Cos[wm*tn-(ph2-ph1)/2];    (* Theoretical cg plot *)
```

Bϕ animations were generated using the above code and substituting the following known Bϕ relations

```
h=y*(-(u*r)-I)*u/r^2*Exp[I*u*r];     (* FT [Signal] x G *)
ph1=k1*r-ArcCos[-k1*r/Sqrt[1+(k1*r)^2]];     (* Bϕ phase relation for f1 *)
ph2=k2*r-ArcCos[-k2*r/Sqrt[1+(k2*r)^2]];     (* Bϕ phase relation for f2 *)
Md=2*Abs[(-(wc/c*r)-I)]*wc/c/r^2*Cos[wm*tn-(ph2-ph1)/2];     (* Theoretical cg plot *)
```

To check the simulator, the AM signal was also applied to a light propagating system. The results yielded a 300 MHz carrier, 20 MHz modulated AM signal animation which linearly increased its phase shift as expected. This was verified by substituting the following known relations:

```
h=y*Exp[I*u*r];    (* FT [Signal] x G *)
Md=2*Cos[wm*tn-( k2*r - k1*r)/2];    (* Theoretical cg plot *)
```

It should be noted that no signal distortion was observed in these simulations. This can be attributed to the fact that the phase and amplitude vs. frequency curves are approximately linear over the bandwidth of the signal ($\Delta f/f_c$ = 40/300 = 1/7.5). The use of linearity constraint can be seen to be justified by plotting $u^2/A^2$ for each field component and noting that it is much less than one over the bandwidth of the signal (Eq. 13 - 15).

$E_r$ ($u^2/A^2$ vs. r)

**Fig. 15**

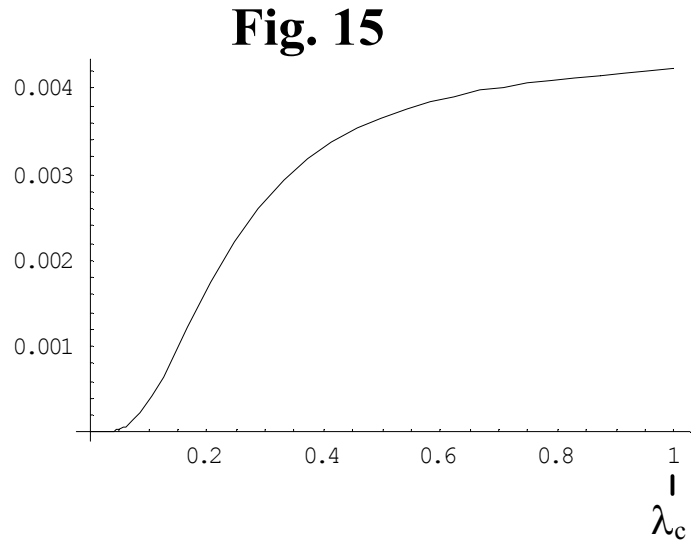

$E_\theta$ ($u^2/A^2$ vs. r)

**Fig. 16**

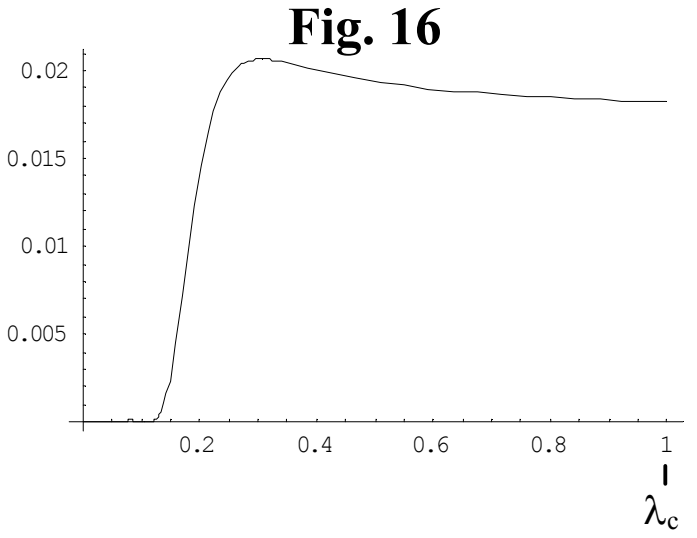

$B_\varphi$ ($u^2/A^2$ vs. r)

**Fig. 17**

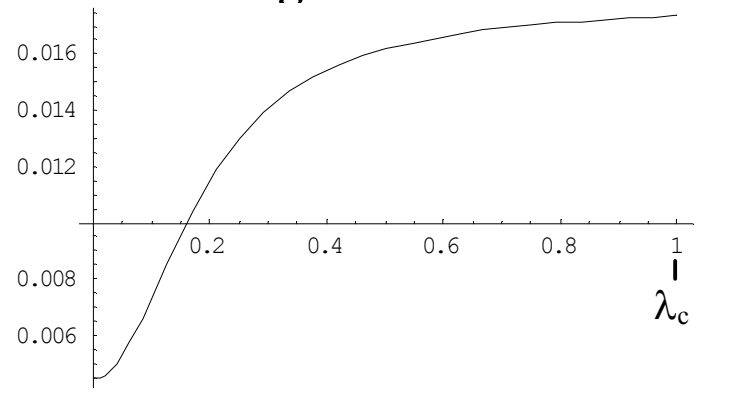

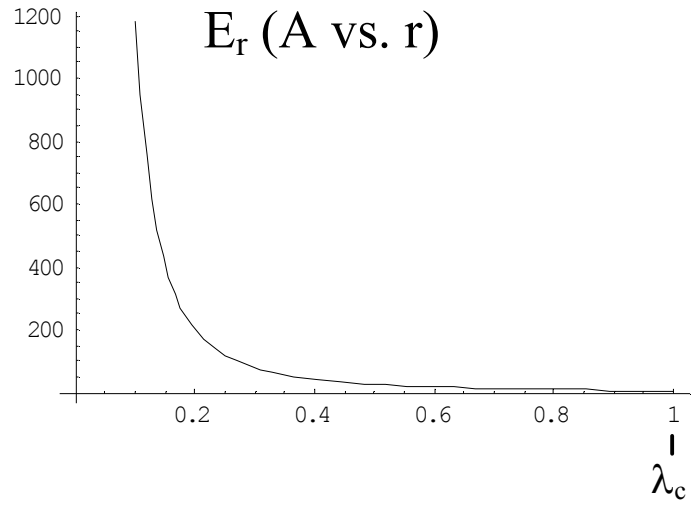

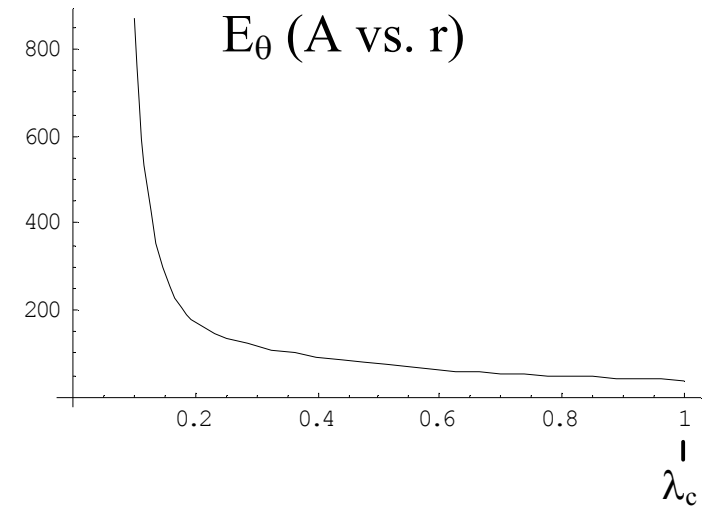

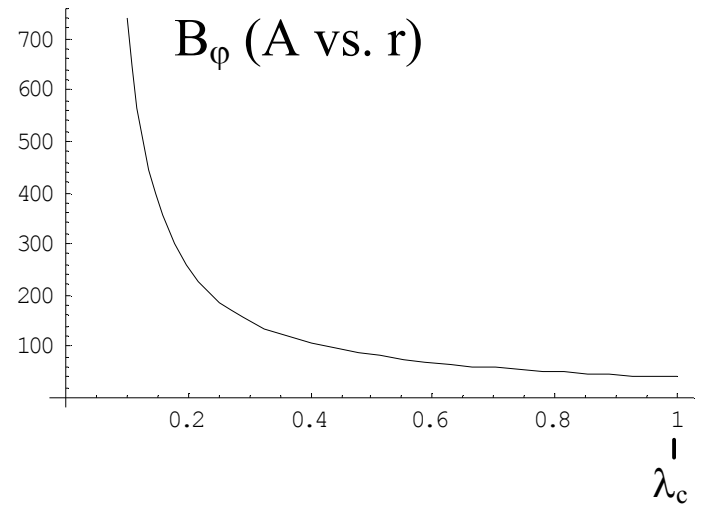

Because of the excellent match between the numerical and theoretical methods, the validity of both methods is confirmed in analyzing the propagation of simple signals in this system. Whereas the theoretical method enables the propagation of simple signals to be clearly understood, the numerical solution is not only useful in verifying the theoretical results, it can also be useful in understanding the propagation of more complex signals which may be difficult to analyze mathematically.

## Experimental verification of $E_\theta$ solution

A simple experimental setup using two dipole antennas, a 437 MHz (68.65 cm wavelength), 2 watt sinusoidal transmitter, and a 500MHz digital oscilloscope has been developed to verify qualitatively the transverse electric field phase vs. distance plot predicted from standard EM theory (Eq. 20, Fig. 5). The phase shift of the received antenna signal (Rx) relative to the transmitted signal (Tx) was measured with an oscilloscope as the distance between the antennas ($r_{el}$) was changed from 5 cm to 70 cm in increments of 5 cm. The data was then curve fit with a 3rd order polynomial. The phase speed vs. distance curve was then generated by differentiating the resultant curve fit equation with respect to space and using (Eq. 8). The group speed vs. distance curve was generated by using the differential relation:

$\Delta\theta = \Delta kr \dfrac{\partial\theta}{\partial kr}$ and inserting it into the relation (Eq. 12): $c_g = \dfrac{\Delta\omega}{\dfrac{\partial\Delta\theta}{\partial r}}$ where: $\Delta\omega = \Delta k\, c$

yielding: $$c_g = \frac{360\, c_o}{\left[ r_{el} \dfrac{\partial^2\theta}{\partial r_{el}^{\,2}} + \dfrac{\partial\theta}{\partial r_{el}} \right]} \qquad \text{where: } r_{el} = \frac{r}{\lambda} \qquad (26)$$

Experimental setup

**Fig. 18**

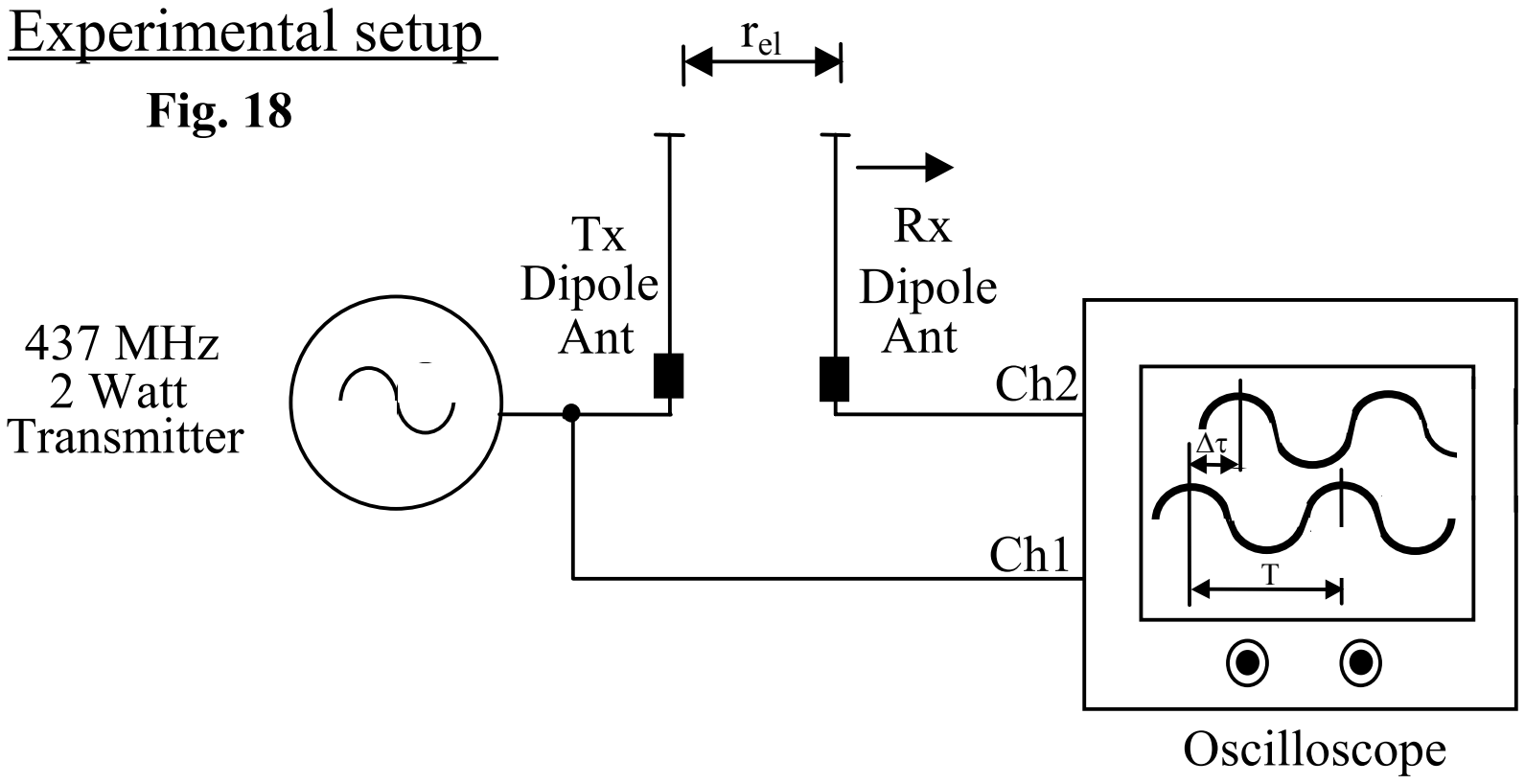

Experimental data

**Fig. 19**

| Data # | $r_{el}$ (cm) | Ph (Deg) |
|---|---|---|
| 0 | 0 | 140.0 |
| 1 | 5 | 111.7 |
| 2 | 10 | 102.4 |
| 3 | 15 | 108.6 |
| 4 | 20 | 121.0 |
| 5 | 25 | 136.6 |
| 6 | 30 | 155.2 |
| 7 | 35 | 170.7 |
| 8 | 40 | 195.5 |
| 9 | 45 | 211.0 |
| 10 | 50 | 245.2 |
| 11 | 55 | 282.4 |
| 12 | 60 | 316.6 |
| 13 | 65 | 325.9 |
| 14 | 70 | 366.2 |

$E_\theta$ phase plot similar to Fig. 5

**Fig. 20**

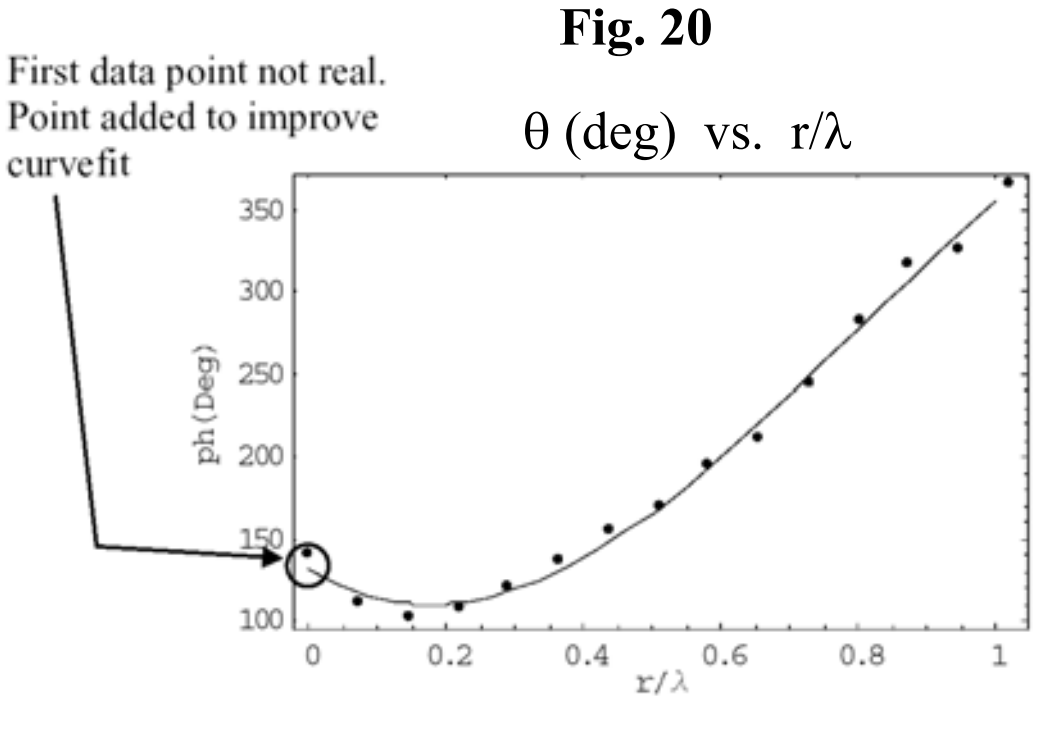

$E_\theta$ curve fit equation

$$ph = (132.2) + (-262.5)r_{el} + (838.9)r_{el}^2 + (-353.4)r_{el}^3 \qquad (27)$$

Resultant $E_\theta$ phase speed and group speed plots - very similar to the predicted plots (Fig. 6, 7)

$c_{ph}/c_o$ vs. $r/\lambda$

**Fig. 21**

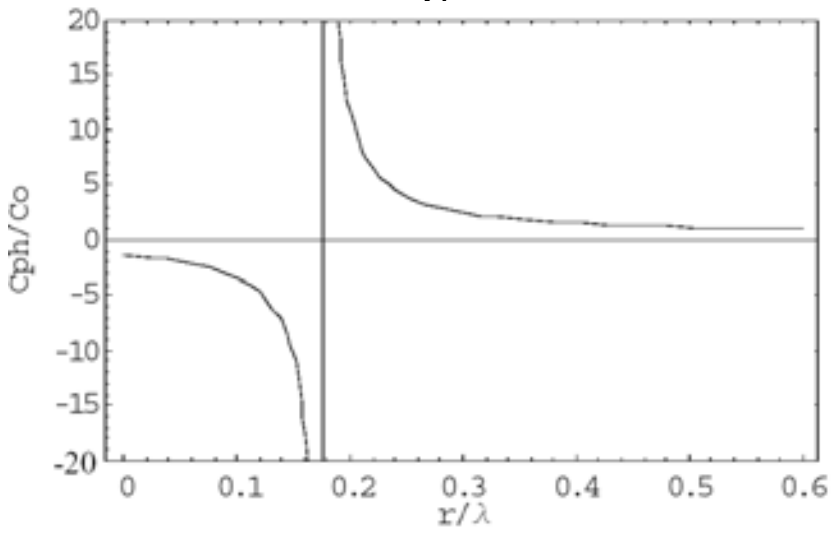

$c_g/c_o$ vs. $r/\lambda$

**Fig. 22**

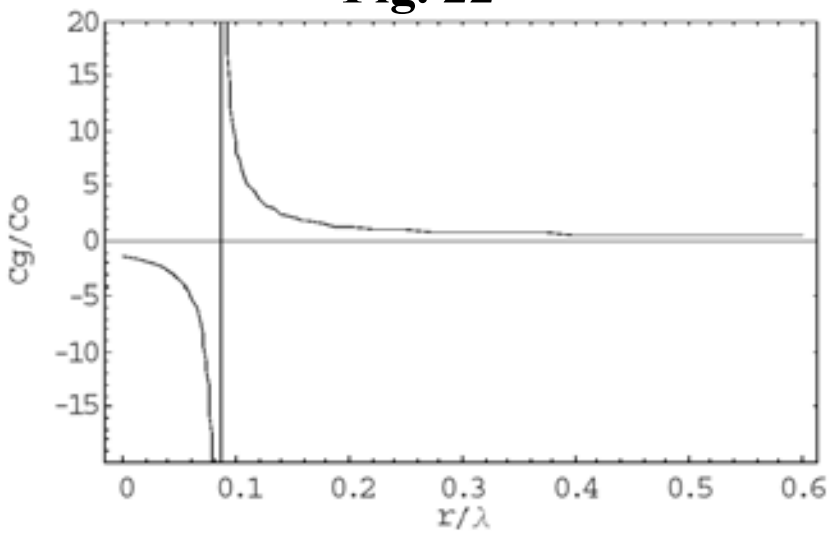

It should be noted that the results are only qualitative due to reflections. Refer to a previous paper written by the author for more details [8].

## Other wave propagation systems with similar superluminal behavior

### Magnetic dipole

Theoretical analysis of a magnetic dipole reveals that the system is governed by the same partial differential equation as the electric dipole with the E and B fields interchanged [2, 4]. The resulting fields are found to be the same as the fields generated by an electric dipole (Eq. 4) and therefore the phase speed and group speed of these fields are the same as (Eq. 18 - 25), except that the E and B fields are interchanged.

### Electric and magnetic quadrapole

Using the same method of analysis as was done for the electric dipole, the scalar potential for an electric quadrapole is found to be [3]:

$$V = -N\left( i\left(\frac{1}{kr} - \frac{3}{(kr)^3}\right) - \frac{3}{(kr)^2} \right)\left(3Cos^2\theta - 1\right)e^{i(kr-\omega t)} \quad (28)$$

The fields can then be calculated using the following relations [6]:

yielding: $$B = \frac{\omega}{c^2}(r \times \nabla V) \qquad E = \frac{ic^2}{\omega}(\nabla \times B) \quad (29)$$

$$B_\phi = \frac{6N\omega}{c^2}\left[\left(\frac{1}{kr} - \frac{3}{(kr)^3}\right)i - \frac{3}{(kr)^2}\right]\left[Cos(\theta)Sin(\theta)\right]\ e^{i(kr-\omega t)} \quad (30)$$

$$E_r = 6Nk\left[\left(\frac{-3}{(kr)^3}\right)i - \frac{1}{(kr)^2} + \frac{3}{(kr)^4}\right]\left[3Cos^2(\theta) - 1\right]\ e^{i(kr-\omega t)} \quad (31)$$

$$E_\theta = 6Nk\left[\left(\left(\frac{1}{kr}\right) - \frac{6}{(kr)^3}\right)i + \frac{6}{(kr)^4} - \frac{3}{(kr)^2}\right]\left[Cos(\theta)Sin(\theta)\right]e^{i(kr-\omega t)} \quad (32)$$

where: $N = \frac{qs^2k^3}{4\pi\varepsilon_o}$ in all the above solutions.

s = Dipole length, q= Charge
k = Wave number, $\varepsilon_o$ = Permitivity

The fields for a magnetic quadrapole are also the same, with the E and B fields interchanged. The phase and group speed of these fields can be determined using the relations (Eq. 8, 12). These results are presented in the next section (Eq. 39 - 47).

## Gravitational quadrapole

For weak and slowly varying gravitational fields, Einstein's equation becomes linearized and reduces to [19, 20]:

$$\nabla^2 V - \frac{1}{c^2}\frac{\partial^2 V}{\partial t^2} = 4\pi G\rho \tag{33}$$

| Where: | | |
|---|---|---|
| | $\rho$ = Mass density | $\nabla^2$ = Laplacian |
| | V = Gravitational potential | c = Speed of light |
| | G = Gravitational constant | t = Time |

Except for the source term, the partial differential equation of the potential is the same as that of an oscillating charge (Eq. 1). Because of this similarity one can then use the oscillating charge solutions by simply substituting: $\varepsilon_o = -1/(4\pi G)$. In addition, because momentum is conserved, a moving mass must push off another mass. The gravitational field generated by the secondary mass adds to the gravitational fields generated by the moving mass, resulting in a linear quadrapole source. The gravitational fields generated by an oscillating mass are therefore of the same form as the fields generated by an electric quadrapole (Eq. 28 - 32).

$$V = -N\left(i\left(\frac{1}{kr} - \frac{3}{(kr)^3}\right) - \frac{3}{(kr)^2}\right)\left(3Cos^2\theta - 1\right)e^{i(kr-\omega t)} \tag{34}$$

It is known that for weak and slowly varying gravitational fields, General Relativity theory reduces to a form of Maxwell's equations [19]. The fields can then be calculated using the following relations:

$$B = \frac{\omega}{c^2}(r \times \nabla V) \qquad E = \frac{ic^2}{\omega}(\nabla \times B) \tag{35}$$

where (E) is the gravitational force vector and (B) is the solenoidal gravitational force vector. The constant (N) can be determined by substituting the relations: $\varepsilon_o = -1/(4\pi G)$ and q = m into the value of (N) used in the electric quadrapole. In addition, this result can be checked by looking at the static quadrapole solution and comparing it to the above solutions in the limit ($kr \to 0$). The results yield:

$$B_\phi = \frac{6N\omega}{c^2}\left[\left(\frac{1}{kr} - \frac{3}{(kr)^3}\right)i - \frac{3}{(kr)^2}\right]\left[Cos(\theta)Sin(\theta)\right]e^{i(kr-\omega t)} \tag{36}$$

$$E_r = 6Nk\left[\left(\frac{-3}{(kr)^3}\right)i - \frac{1}{(kr)^2} + \frac{3}{(kr)^4}\right]\left[3Cos^2(\theta) - 1\right]e^{i(kr-\omega t)} \tag{37}$$

$$E_\theta = 6Nk\left[\left(\left(\frac{1}{kr}\right) - \frac{6}{(kr)^3}\right)i + \frac{6}{(kr)^4} - \frac{3}{(kr)^2}\right]\left[Cos(\theta)Sin(\theta)\right]e^{i(kr-\omega t)} \tag{38}$$

where: $N = -G\, m\, s^2\, k^3$, G = Grav const., m = mass, s = Dipole length, k = Wave number

The phase and group speed relations for these fields can then be determined by using the phase and group speed equations derived earlier in the paper (Eq. 8, 12). It should be noted that all of the plots look very similar to those of an electric dipole (see page 4).

**$B_\phi$ phase, phase speed, group speed analysis**

$$ph = kr - ArcTan\left[\frac{kr}{3} - \frac{3}{kr}\right] \underset{kr \ll 1}{\approx} \frac{\pi}{2} + \frac{1}{45}(kr)^5 + O(kr)^7 \tag{39}$$

$$c_{ph} = c_o\left(1 + \frac{3}{(kr)^2} + \frac{9}{(kr)^4}\right) \tag{40}$$

$$c_g = \frac{c_o\left[3(kr)^2 + (kr)^4 + 9\right]^2}{(kr)^4\left[45 + 9(kr)^2 + (kr)^4\right]} \tag{41}$$

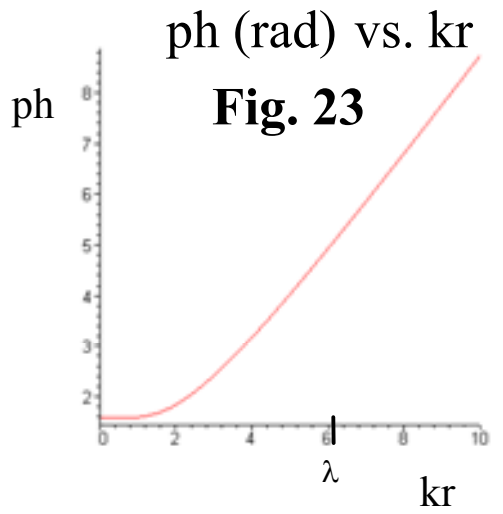

ph (rad) vs. kr
**Fig. 23**

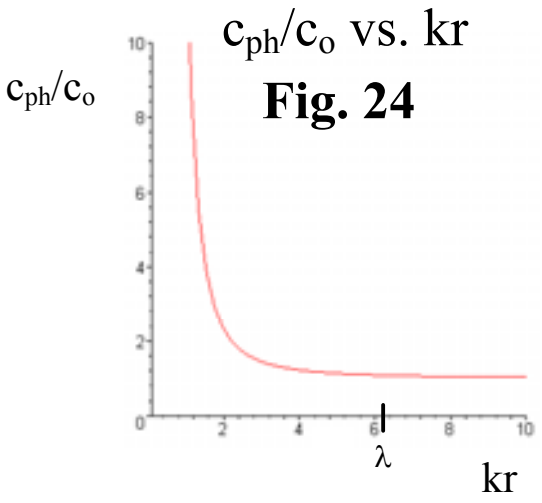

$c_{ph}/c_o$ vs. kr
**Fig. 24**

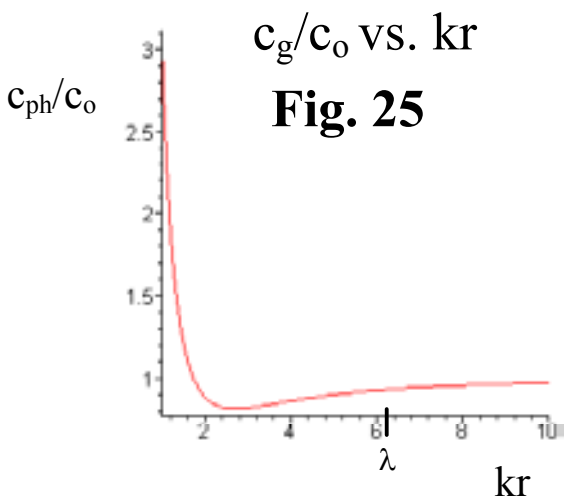

$c_g/c_o$ vs. kr
**Fig. 25**

**Er phase, phase speed, group speed analysis**

$$ph = kr + ArcTan\left[\frac{3}{kr - \frac{3}{kr}}\right] \underset{kr \ll 1}{\approx} \frac{1}{45}(kr)^5 + O(kr)^7 \tag{42}$$

$$c_{ph} = c_o\left(1 + \frac{3}{(kr)^2} + \frac{9}{(kr)^4}\right) \tag{43}$$

$$c_g = \frac{c_o\left[(\mathrm{kr})^4 + 3(\mathrm{kr})^2 + 9\right]^2}{(\mathrm{kr})^4\left[(\mathrm{kr})^4 + 9(\mathrm{kr})^2 + 45\right]} \tag{44}$$

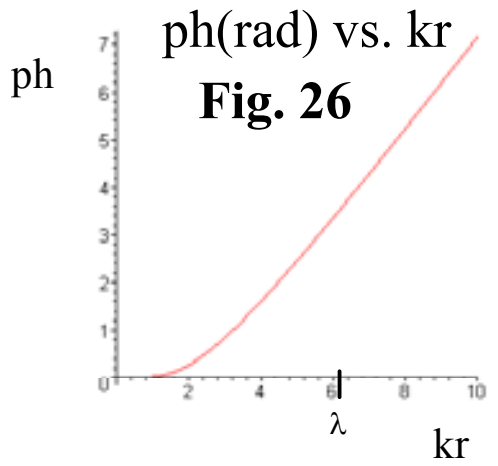

ph(rad) vs. kr
**Fig. 26**

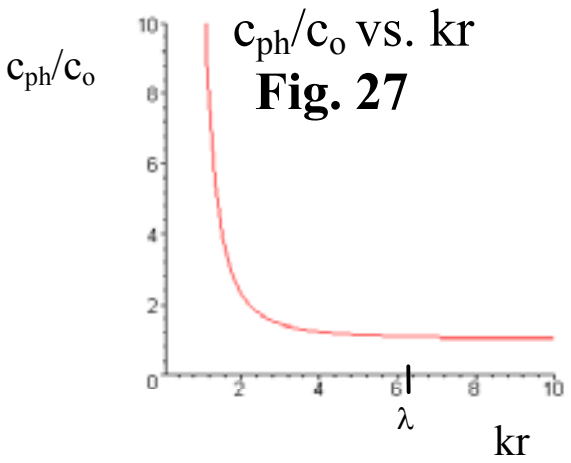

$c_{ph}/c_o$ vs. kr
**Fig. 27**

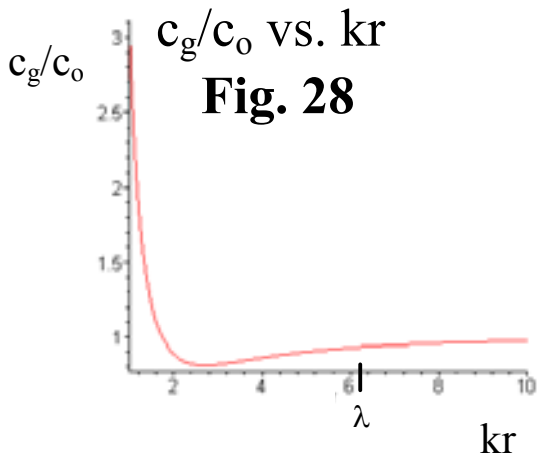

$c_g/c_o$ vs. kr
**Fig. 28**

**$E_\theta$ phase, phase speed, group speed analysis**

$$ph = kr + ArcTan\left[\frac{\frac{-6}{(kr)^3} + \frac{1}{kr}}{\frac{6}{(kr)^4} - \frac{3}{(kr)^2}}\right] \underset{kr \ll 1}{\approx} -\frac{1}{30}(kr)^5 + O(kr)^7 \tag{45}$$

$$c_{ph} = c_o\left(\frac{36 - 3(kr)^4 + (kr)^6}{(kr)^6 - 6(kr)^4}\right) \tag{46}$$

$$c_g = \frac{c_o\left[36 - 3(kr)^4 + (kr)^6\right]^2}{(kr)^{12} - 3(kr)^{10} + 18(kr)^8 + 252(kr)^6 - 1080(kr)^4} \tag{47}$$

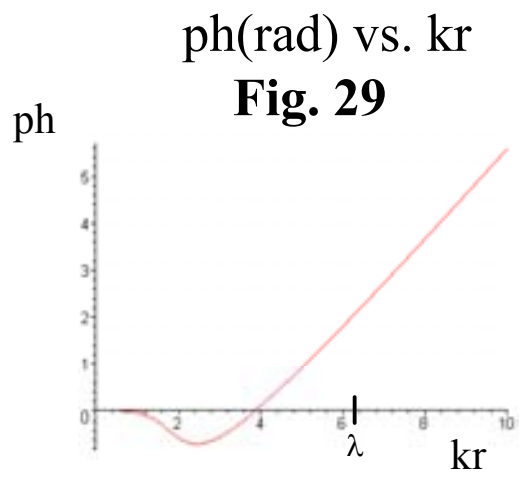

ph(rad) vs. kr
**Fig. 29**

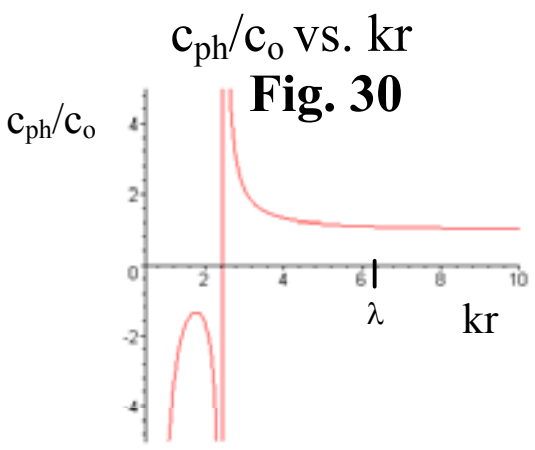

$c_{ph}/c_o$ vs. kr
**Fig. 30**

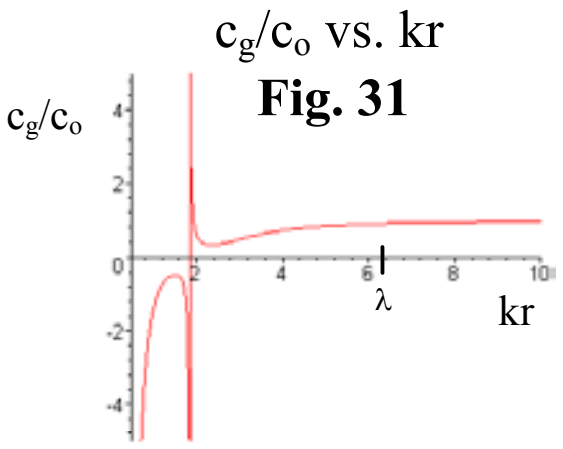

$c_g/c_o$ vs. kr
**Fig. 31**

**Field contour plots** (linear quadrapole in center and vertical)

Contour plots of the fields (Eq. 36 - 38) using Mathematica software yields [9]:

$B_\phi$ contour plot

**Fig. 32**

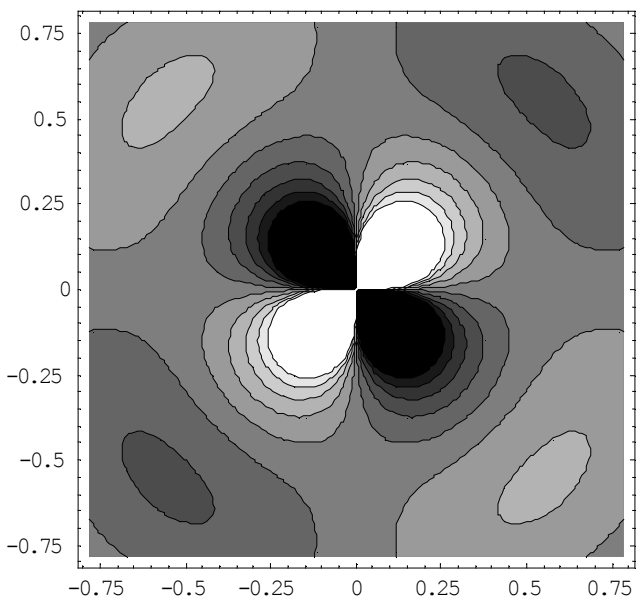

$E_r$ contour plot

**Fig. 33**

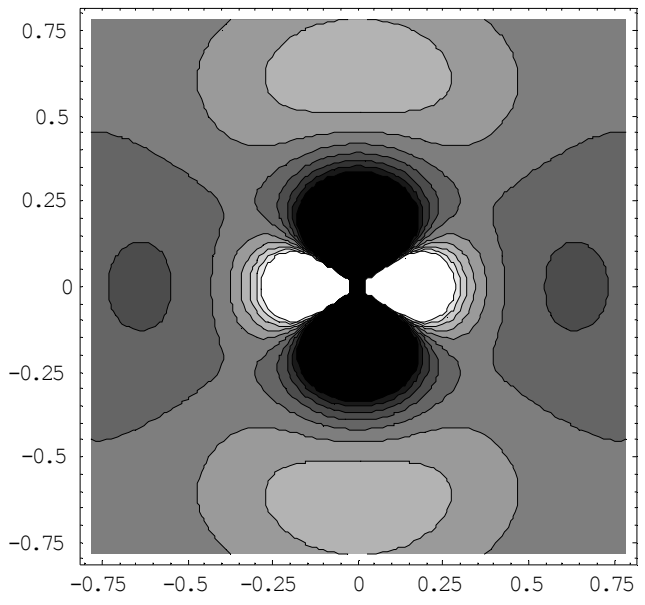

$E_\theta$ contour plot

**Fig. 34**

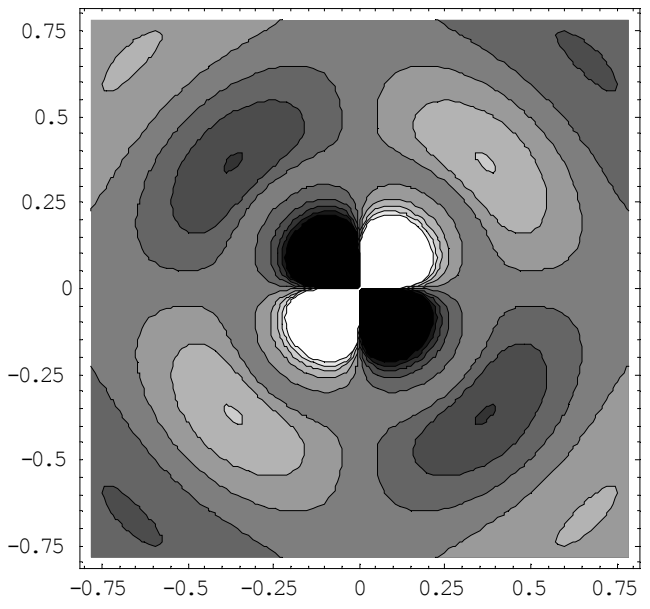

**Total E field plots** (linear quadrapole in center and vertical unless specified)

Using vector field plot graphics in Mathematica software [9], the Er and $E_\theta$ can be combined and plotted as vectors (Fig. 35). A more detailed plot of the total E field can be obtained by using the fact that a line element crossed with the electric field = 0. A contour plot of the resulting relation yields the total E field plots below (Fig. 36 - 37).

Near-field
Vector plot

**Fig. 35**

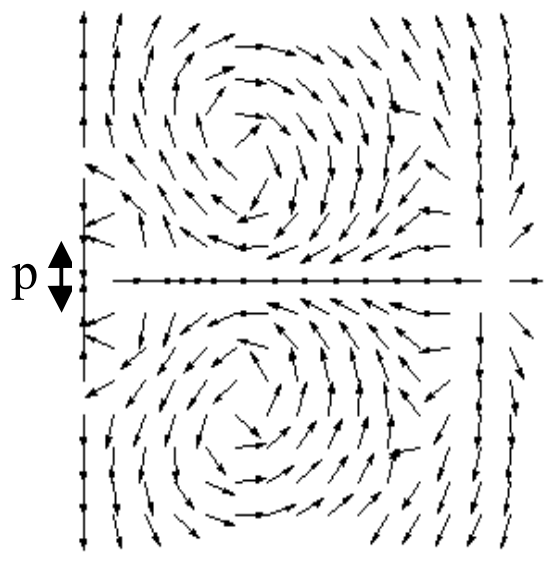

Near-field
Contour plot

**Fig. 36**

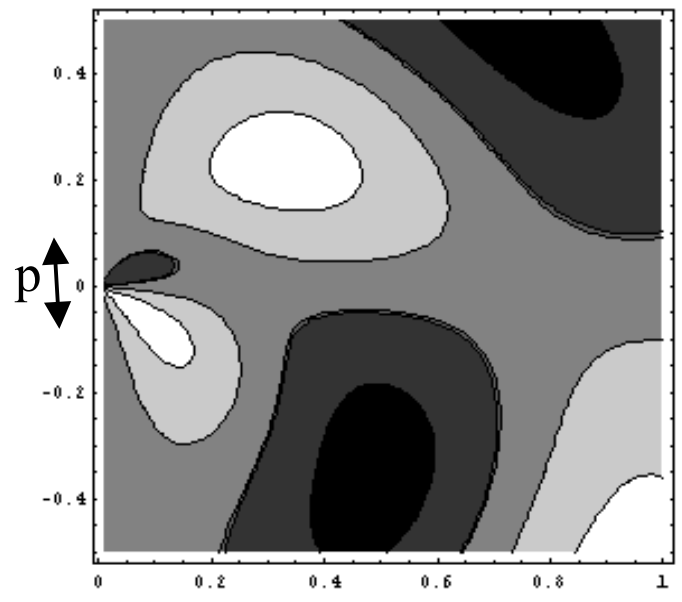

Far-field
Contour plot

**Fig. 37**

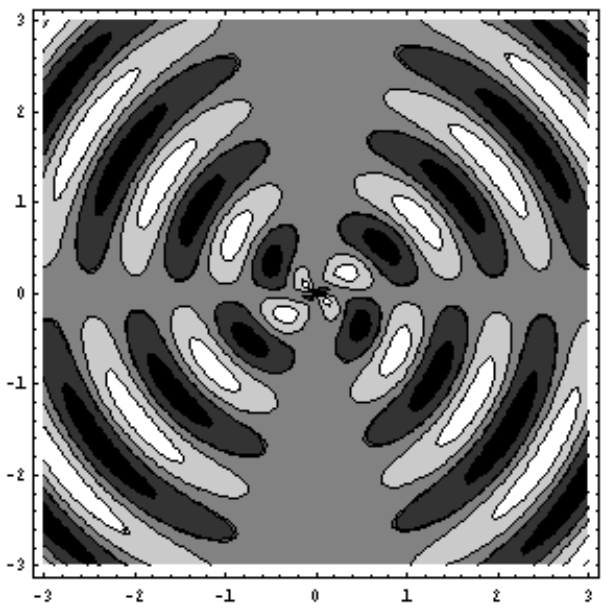

**Experimental evidence of superluminal gravitational fields**

Evidence of infinite gravitational phase speed at zero frequency has been observed by a few researchers by noting the high stability of the earth's orbit about the sun [21, 22]. Light from the sun is not observed to be collinear with the sun's gravitational force. Astronomical studies indicate that the earth's acceleration is towards the gravitational center of the sun even though it is moving around the sun, whereas light from the sun is observed to be aberated. If the gravitational force between the sun and the earth were aberated then gravitational forces tangential to the earth's orbit would result, causing the earth to spiral away from the sun, due to conservation of angular momentum. Current astronomical observations estimate the phase speed of gravity to be greater than $2 \times 10^{10}$c. Arguments against the superluminal interpretation have appeared in the literature [23, 24]

**Information speed**

If an amplitude-modulated signal propagates a distance (d) in time (t), then the information contained in the modulation propagates at a speed:

$$c_{inf} = d/(t+T) \tag{48}$$

where (T) is the amount of time the modulated signal must pass by the detector in order for the information to be determined. The information in the wave is determined by measuring the amplitude, frequency, and phase of the wave modulation envelope.

If a wave is propagated across distances in the farfield of the source, then the wave information speed is approximately the same as the wave group speed. This is because the wave propagation time (t) is much greater than the wave information scanning time (T), consequently: $c_{inf} = d/t = c_g$ .

In the nearfield of the source, if nothing is known about the type of modulation, then the scanning time (T) can be much larger than the wave propagation time (t), thereby making the wave information speed much less than the wave group speed. This can be understood by noting that several modulation cycles are required for a Fourier analyzer to be able to determine the wave modulation amplitude, frequency, and phase. But if the type of modulation is known, then only a few points of the modulated signal need to be sampled by a detector in order to curve fit the signal and therefore determine the modulation information. If the noise in the signal is very small then the signal scanning time (T) can be made much smaller than the signal propagation time (t), consequently: $c_{inf} \sim d/t = c_g$ .

**Relativistic consequences**

According to the relativistic Lorentz time transform (Eq. 49), if an information signal can propagate at a speed (w) faster than the speed of light (c), then the signal can be reflected by a moving frame (v) located a distance (L) away and the signal will arrive before the signal was transmitted ($\Delta t' < 0$). Since the information in the signal can be used to prevent the signal from being transmitted, this results in a logical contradiction (violation of causality). How can the signal be detected if it was never transmitted? Consequently, Einstein in 1907 stated that superluminal signal velocities are incompatible with Relativity theory [25].

**Fig. 38**

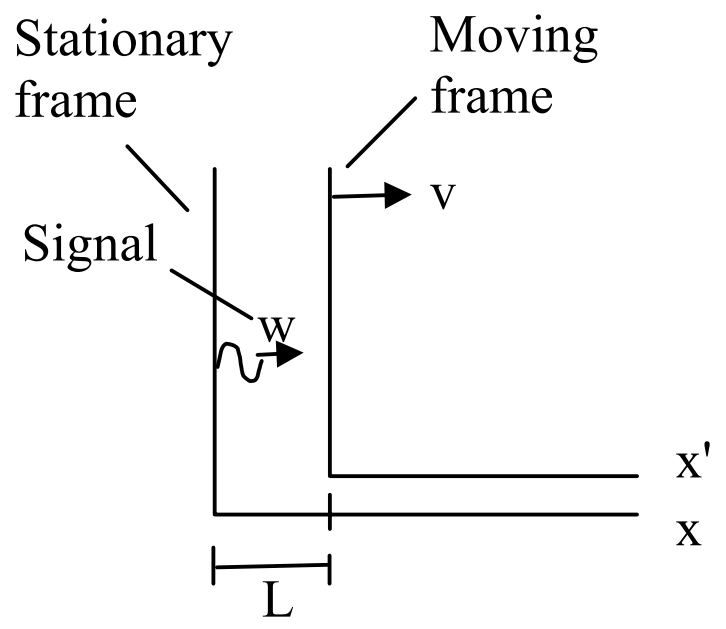

$$\Delta t' = \gamma\left(\Delta t - \frac{v}{c^2}\Delta x\right) \underset{w > \frac{c^2}{v}}{\Rightarrow} \text{Neg} \tag{49}$$

where: $\Delta t = \frac{L}{w} \qquad \Delta x = L$

$$\gamma = \frac{1}{\sqrt{1 - \frac{v^2}{c^2}}}$$

Because Relativity theory predicts that a moving reflector (which has mass) can never move faster than light ($v < c$), then in order for ($\Delta t' > 0$) the signal propagation speed must be less than the speed of light ($w<c$).

## Argument against the superluminal interpretation

Some physicists have proposed that a dipole source generates position, velocity, and acceleration-dependent propagating fields, each of which propagate at the speed of light [5, 26]. It is argued that the interference of these field components gives the illusion that they propagate superluminally. Although this is a plausible explanation it is not clear that this is true. It can be shown using Maxwell's equations that the fields generated by a dipole source can be modeled by two coupled partial differential equations in terms of the scalar potential (V) and the vector potential (A):

$$-\nabla^2 V - \frac{\partial}{\partial t}(\nabla \cdot A) = \frac{\rho}{\varepsilon_o} \quad \text{and} \quad -c^2\nabla^2 A + c^2\nabla(\nabla \cdot A) + \frac{\partial}{\partial t}\nabla V + \frac{\partial^2 A}{\partial t^2} = \frac{j}{\varepsilon_o} \qquad (50)$$

where (j) is the source current density, (ρ) is the source charge density, and (c) is the speed of light [5(Ch 18-6), 3(Ch 6.4)]. In order to simplify the equations, an equation is introduced with no physical basis in EM theory (gauge condition). The most common is the Lorentz gauge: $\nabla \cdot A = -\frac{1}{c^2}\frac{\partial V}{\partial t}$ which yields the following simple decoupled 2nd order partial differential equations:

$$\nabla^2 A - \frac{1}{c^2}\frac{\partial^2 A}{\partial t^2} = \frac{j}{\varepsilon_o c^2} \quad \text{and} \quad \nabla^2 V - \frac{1}{c^2}\frac{\partial^2 V}{\partial t^2} = \frac{-\rho}{\varepsilon_o} \qquad (51)$$

$$\text{yielding solutions:} \quad V = \frac{1}{4\pi\varepsilon_o}\int_{Vol}\frac{\rho}{r}\,dVol \quad \text{and} \quad A = \frac{\mu_o}{4\pi}\int_{Vol}\frac{j}{r}\,dVol \qquad (52)$$

where the charges and currents in the solutions are evaluated at a previous time (retarded): t - r/c. It is argued here that the Lorentz gauge is not the only gauge that can be used to simplify the two PDEs. Many other gauges are possible including the Coulomb gauge ($\nabla \cdot A = 0$), which in particular is known to yield an instantaneous scalar potential [3(Ch 6.5)]. This shows that although the dipole solution is thought to be gauge independent, the form of the solution may differ when different gauge conditions are used, giving rise to different interpretations of the speed of the resulting propagating fields. Only the PDE's (Eq. 50) are physical in EM theory, and can probably only be solved numerically without using a gauge relation, yielding fields that propagate as shown in this paper (see pages 4-6).

## Conclusion

The analysis presented in this paper has shown that the fields generated by an electric or magnetic dipole or quadrapole, and also the gravitational fields generated by a quadrapole mass source, propagate superluminally in the nearfield of the source and reduce to the speed of light as they propagate into the farfield. The group speed of the waves produced by these systems has also been shown to be superluminal in the nearfield. Although information speed can be less than group speed in the nearfield, it has been shown that if the method of modulation is known and provided the noise of the signal is small enough, the information can be extracted in a time period much smaller than the wave propagation time. This would therefore result in information speeds only slightly less than the group speed which has been shown to be superluminal in the nearfield of the source. It has also been shown that Relativity theory predicts that if an information signal can be propagated superluminally, then it can be reflected by a moving frame and arrive at the source before the information was transmitted, thereby enabling causality to be violated.

Given these results, it is at present unclear how to resolve this dilemma. Relativity theory could be incorrect, or perhaps it is correct and information can be sent backwards in time. Perhaps as suggested by the 'Hawking chronology protection conjecture' [27], nature will intervene in any attempt to use the information to change the past. Therefore information can be propagated backwards in time but it cannot be used to change the past, thereby preserving causality. Another possibility is that according to the 'many-worlds' interpretation of quantum mechanics [28], multiple universes are created any time an event with several possible outcomes takes place. If this interpretation is correct, then information can be transmitted into the past of alternative universes, thereby preserving the past of the universe from which the signal was transmitted.

In addition to the theoretical implications of the research discussed above, it may also have practical applications, such as increasing the speed of electronic systems that will soon be limited by light-speed-time delays. It should also be possible to reduce the time delays inherent in current astronomical observations by monitoring lower frequency EM and eventually gravitational radiation from these sources. Lastly, using low frequency EM transmissions, it should be possible to reduce the long communication time delays to spacecraft.